# Mapping the network structure of science parks: An exploratory study of cross-sectoral interactions reflected on the web

David Minguillo and Mike Thelwall
*Statistical Cybermetrics Research Group, SCSIT, University of Wolverhampton*

**Abstract**
**Purpose** - This study introduces a method based on link analysis to investigate the structure of the R&D support infrastructure associated with science parks in order to determine whether this webometric approach gives plausible results.
**Design/Methodology/Approach** - Three science parks from Yorkshire and the Humber in the UK were analysed with webometric and social network analysis techniques. Interlinking networks were generated through the combination of two different data sets extracted from three sources (Yahoo!, Bing, SocSciBot).
**Findings** - These networks suggest that institutional sectors, representing business, universities and public bodies, are primarily tied together by a core formed by research institutions, support structure organisations and business developers. The comparison of the findings with traditional indicators suggests that the web-based networks reflect the offline conditions and policy measures adopted in the region, giving some evidence that the webometric approach is plausible to investigating science park networks.
**Originality/value** – This is the first study that applies a web-based approach to investigate to what extent the science parks facilitate a closer interaction between the heterogeneous organisations that converge in R&D networks. This indicates that link analysis may help to get a first insight into the organisation of the R&D support infrastructure provided by science parks.
**Paper Type** Research paper

## Introduction

The complexity and uncertainty introduced by a global and highly-competitive economy has led to the design of new models of production and economic development based on knowledge-intensive high technology industries (Castells, 2000). Consequently, most developed countries have increasingly adopted research and development (R&D) programs to promote and strengthen the collaboration between research institutions and companies in order to exploit the accumulated science and technology research. Some of these initiatives and important investments have materialised through the creation of physical spaces, such as science-, research- or technology parks and high technology clusters, which provide a micro-environment that agglomerates and interconnects a wide range of interdependent organisations belonging to the public, private and academic sectors (Gordon & McCann, 2000). These artificial environments offer the capability and resources to facilitate the transfer and application of knowledge, which in turn, promotes the development and commercialisation of innovative services and products (Breschi & Catalini, 2010; Etzkowitz, 2008). The science parks (SPs) started to appear in most developing countries in the 1980s and have experienced a rapid adoption among emerging economies. Policy makers increasingly consider science parks to be a key tool for raising the level of technological sophistication and reactivating local economies, thus moving towards the development of a knowledge intensive economy (Cumbers & Mackinnon, 2004; Porter, 2000). In the UK, the park building boom started in the 80s due to three principal causes; first, a strong investment in the construction of the parks and a reduction in research funding, undermining the very strengths which SPs sought to exploit. Second, as a policy initiative from the government to incentive the industrial resurgence and create job opportunities and thus overcome the severe recession and unemployment of 79-81 (Quintas, et al., 1992), and third, as a result of the transformation from polytechnic institutions into universities (Siegel, Westhead, & Wright, 2003a).

SPs are seen as micro-communities of cross-sectoral and interdependent organisations united by an identity, a mission, a set of routines and a strategic core, which interacts with the external environment as a unified entity that changes over time and that should be studied from a social network perspective (Bøllingtoft & Ulhøi, 2005;

Phan, Siegel, & Wright, 2005). These R&D networks have an institutional intermediary role, which gives them an inherent multidimensional and synergetic nature, encompassing different kinds of organisations and sectors which can be studied at different levels of analysis (Tijssen, 1998). However, there is currently no data available nor a systematic framework to study the diversity of linkages established within and across industry, higher education, and government agents involved in SPs (Phan, Siegel, & Wright, 2005). This makes it difficult to understand and assess the relational and intermediary capacity of these organisational innovation structures, which aim to strengthen the cross-sectoral connections that are necessary to generate highly innovative environments. However, we agree with Castells that cyberspace enables the analysis of different institutional networks that coexist in the digital dimension and become interconnected through hyperlinks, thus building global digital networks of interactions, which transcend territorial and institutional boundaries (2009:4, 24). Therefore, we believe that the important role of ICTs and the Web for these agglomerations (Steinfield & Scupola, 2008) could help tackle the complexity of these diversified R&D networks and make it possible to deal with the relational structure generated by heterogeneous groups of organisations, which otherwise would need to be studied through a set of widely accepted indicators.

This fact makes the emerging research area webometrics and especially the link analysis technique an interesting method to study the configuration and dynamics of SPs, particularly as intermediaries and connectors of different spheres. Hyperlinks are studied in fields such as Physics, Computer science and Information science. In the last field, the analogy between citation networks and collections of hyperlinked documents draws special attention to the behavioural foundations of hyperlinks (Ackland, 2009; Thelwall, 2006). Information scientists usually also acquire a social science perspective, viewing the Web as a complex and multilayered relational space which enables various forms of social, economic, and political behaviour. Here the focus is on studying the underlying structure and value of networks formed by people and organisations which might reveal offline phenomena (Bar-Ilan, 2005; Park & Thelwall, 2003; Thelwall, 2004; Thelwall, et al., 2005). Encouraging results in the study of academic institutions (Aguillo, et al., 2006; Thelwall, 2001) have led to a wide range of application contexts related to the business performance and have shown that R&D investments and revenues may related to web visibility as long as the organisations are homogeneous (Vaughan, 2004; Vaughan & Wu, 2004) and their size are not considered (Martínez-Ruiz & Thelwall, 2010). Indirect links have also been used to map businesses market positions (Vaughan & You, 2006), and linking pages have been analysed to find out the motivations for link creation (Vaughan, Kipp, & Gao, 2007). Similar approaches have been applied to the international banking industry (Vaughan & Romero-Frias, 2010), political communication (Park & Thelwall, 2008), local and national government bodies (Holmberg, 2010; Petricek, et al., 2006), and the structural dynamics between organisations in a region (Faba-Pérez, et al., 2004).

Nevertheless, few studies have focused on studying different aspects of R&D networks on the web. The first attempt examined the feasibility of webometrics to explore the virtual social structure and to measure the academia-government-industry dynamics (Boudourides, et al., 1999). Later, Leydesdorff and Curran (2000) used word co-occurrence and hypertextual links to conclude that the relationship between the university and the other two sectors is stronger at the international level while national economies promote stronger industry-government linkages. The study of Stuart and Thelwall (2006) analysed the intra- and inter-sector collaboration in the automobile industry at regional level in the UK, finding that the lack of web connectivity only partially reflects the offline collaboration. Similarly, another study tried to trace the national communicative change of the Triple Helix in Spain by means of outlinks and co-outlinks (Garcia-Santiago & de Moya-Anegon, 2009). However, the main weakness of these studies might be the broad context of study and the subjective criteria to select the different organisations to be involved in an innovation system, especially in a not codified context such as the Web. To our knowledge, only Ortega's study (2003) examined the science and technology interactions in a specialized environment, using the outlinks of research centres affiliated to two Bio-related institutes and showed that the different centres occupied positions related to government, academia, or industry according to their activities.

Due to the low number of studies and questionable results, web-based findings are merely suggestive of hypotheses that need further investigation, being necessary to develop a set of methods for studying the online nature of dynamic and innovative environments. It may allow taking advantage of the Web to provide a broad overview of the complex configuration of R&D networks, which could then be further analysed and disclosed by other approaches. As governmental policy tools designed to develop support infrastructures that promote a close institutional collaboration, the purpose of SPs is to facilitate the exploitation of a research base which may lead to socio-economic growth, and therefore it could be useful to investigate whether hyper-links are able to reflect the wide range of R&D linkages. Consequently, this exploratory study has two main objectives: First, to develop a methodology for collecting and analysing web-based interlinking structures to identify and study R&D networks that are supported and tied together by SPs in order to determine to what extent hyperlinks can detect the mutual and diversified interactions that exist within organizational innovation structures, and how these heterogeneous organisations are interconnected; and second, to compare the key features of the web-based structures with offline features that are widely applied to evaluate the R&D conditions in the UK in order to determine the validity and viability of the webometric approach to represent and analyse the mutual interactions among the varied organisations involved in R&D networks. These objectives can be summarised by the following research questions:

- Can webometrics methods identify interactions between institutional sectors as well as interactions between various types of organisations associated with SPs?
- Are the main features of SP interlinking networks in line with findings of official reports and surveys which are essential sources of information to evaluate the R&D infrastructure in the UK?

In addressing these questions, we investigate the viability of link analysis to identify SPs' intermediary role. As key organizational innovation structures, SPs are used to analyse part of the web-based organization of the R&D support infrastructures developed in a given region but this study does not investigate specific R&D aspects of the networks associated to SPs, such as the transfer of knowledge (Siegel, et al., 2003b; Westhead & Storey, 1995), firm performance and formation (Löfsten & Lindelöf, 2002; Quintas et al., 1992) or best management practices (Autio & Klofsten, 1998), which are normally studied trough another proxies. We also wish to determine whether the web-based and offline characteristics reflect similar patterns in order to investigate whether the general picture provided by the web may shed new light on the structures established in these dynamic and innovative environments.

## Methods

The objects of analysis of this study are SPs indexed as full members by the United Kingdom Science Park Association (UKSPA) in the region of Yorkshire and the Humber, and which have their own website and provide a list of their tenants with their respective URLs. This region was selected since three of its four SPs indexed by the UKSPA were considered adequate for the exploratory study, but most importantly because it is an interesting region that has invested in a growing innovative infrastructure as a tool for an industrial restructuring (Huggins & Johnston, 2009). In addition, the SPs in this region offer a wide range of company sizes; from small and medium-sized enterprises and spin-outs to large manufacturing groups, and types; from specialised knowledge and high technology to advanced manufacturing. This heterogeneity is important in order to gain good insights into the different organisations involved in this context, as well as the degree of interaction between the different organisations and sectors.

**Table I. Number of websites linked to by the SPs and their type of relationship with the SPs**

|  | Adv. Man. Park | Leeds Inn. | Yorkshire SP | *Total* |
|---|---|---|---|---|
| External links | 33 | 54 | 123 | 210 |
| Organisations | 32 | 48 | 103 | 183 |
| **Type of relationship** | | | | |
| Tenant | 26 (81%) | 34 (71%) | 81 (79%) | 141 (77%) |
| Information |  | 3 (6%) | 18 (17%) | 20 (11%) |
| Support | 3 (9%) | 4 (8%) |  | 7 (4%) |
| Partnership |  | 5 (10%) | 2 (2%) | 6 (3%) |
| Other | 2 (6%) |  | 1 (1%) | 3 (2%) |
| Membership |  | 2 (4%) | 1 (1%) | 3 (2%) |
| Incubator | 1 (3%) |  |  | 2 (1%) |

To obtain data on the SP link networks, the websites of the *Advanced Manufacturing Park* (attheamp.com), *Leeds Innovation Centre* (leedsinnovationcentre.com), and *York Science Park* (yorksciencepark.co.uk) were crawled with SocSciBot in May 2010. The web crawler identified site outlinks to websites with potentially a formal or informal relationship with the SPs. The crawler collected 215 site outlinks which were then manually checked to identify the type of relationship with the SPs (see Table I). Each targeted website was classified by sector (Industry, Academia, Government), following similar criteria used by Ortega (2003), and according to nine different categories that may be relevant within R&D networks (Etzkowitz, 2008; Howells, 2006; OECD, 2002). The number of relevant websites identified was 190, but after reducing the URLs to their respective domains (e.g., wlv.ac.uk) or sub-domains (e.g., cybermetrics.wlv.ac.uk) to assign each (sub-)domain to an organization and minimise the impact of multiple levels of websites, the number of organizations was reduced to 183 (see Table I). Therefore, the total number of websites analysed (including the SPs) in the study was 186. The problem of websites with multiple domain names were resolved through the selection of the domain name with the highest number of pages, inlinks, outlinks, and the longest time of activity according to Internet Archive.

In order to develop an effective and reliable method to map and analyse the cross-sectoral interactions and relationships established by means of hyperlinks between the websites, the data sets for the analysis were collected from three different sources. The complexity of the web and the low overlap in coverage by search engines makes it necessary to combine various tools and sources to obtain the most reliable results (Thelwall, 2008a). For this reason, the web crawler SocSciBot and the commercial search engines Yahoo! and Bing were all used to gather the data. The networks formed by the direct links between the set of websites linked by each SP reveal bidirectional connections within the sets of websites in each data set. Due to the lack of studies which use different data sets to compare the linkage between peers, it was first necessary to compare the results from the networks based on the in- and outlinks to observe if similar structures are obtained, as well as the differences and similarities to identify which data might provide the best results.

The set of site inlinks (links pointing to a web domain or sub-domain from another domain or sub-domain) were retrieved from the commercial search engine Yahoo! with the help of the free LexiURL Searcher software. Using this software, we were also able to automatically split the queries whose results exceeded the maximum of 1,000 hits permitted when using Yahoo! for this purpose (Thelwall, 2008b). In this way, we obtained up to 19,619 inlinks for each query. The set of site outlinks (links pointing from the domain or sub-domain to another domain or sub-domain) were collected using both the web crawler SocSciBot and the search engine Bing. Combining the web crawler and search engine data was designed to minimise the bias and limitations of each tool and achieve the most complete list of outlinks possible. Due to the time and resources required by the web crawler, each website was crawled to a search depth of two levels only. The crawler made it possible to collect updated information and include sub-domains in the second data set since Bing is not able to retrieve the outlinks of sub-domains. On the other hand, Bing's coverage complemented the relatively superficial crawling when large websites were analysed. The outlinks were collected from Bing through LexiURL searcher and the capability to split the queries with more than 1,000 results was used as well. The comparison and combination of the results was important to improve the quality and quantity of the data set, as illustrated by the low average (4%) overlap among site outlinks.

The number of inlinks collected through Yahoo! was 337,911 and the number of outlinks gathered through SocSciBot and Bing were 6,597 and 104,890 respectively. After the links were reduced to (sub-)domains, the duplicates of each website were eliminated and the outlinks were combined, 183,006 inlinks and 80,588 outlinks remained. This reduction of 46% for the inlinks and 28% for the outlinks suggests that a high proportion of inlinks tend to come from the same group of websites while the outlinks tend to be spread around a wide range of websites. In addition, Yahoo! having 183,006 inlinks and Bing having only 77,657 outlinks suggests that the websites studied do not link to the same extent as they are linked to. Even taking into account that Bing does not allow retrieving outlinks of sub-domains, which was relevant to only 7 (3.7%) websites of the sample with 11,594 (6%) inlinks, the in-data set through Yahoo! still returned 90,824 (113%) more links than the out-data set through Bing. Assuming that the results of the two sources have similar levels of accuracy, this means that the websites receive twice as many links than the ones they give. It is important to note that this comparison reflects two different behaviours (inlinks and outlinks) and not the coverage of both search engines, although the number of in- and outlinks returned could be biased towards the coverage of both search engines.

Next, a matrix of hyperlinks was constructed for each SP and data set. The frequencies with a value greater than or equal to 1 were dichotomised to measure the number of websites that were interconnected rather than the intensity of the connections, and the diagonal entries (selflinks) were eliminated. Dichotomization is recommended because the low frequencies and the direction of the direct links established between a pair of heterogeneous websites may not be representative of offline relationships, leading to misunderstandings. The difference in the link behaviour among various types of websites makes it necessary to only take a large number of direct links into account to identify patterns of the strength and direction of a relationship. Finally, the networks were represented and analysed with the help of Social Network Analysis (SNA) and visualisation techniques. Different cohesion and centrality measures were calculated using UCINET and represented with NetDraw.

Hence, the data used in this study is divided in two different unidirectional data sets, one based on inlinks collected with Yahoo! and another based on outlinks collected with SocSciBot & Bing, which together contain all the links pointing to and from the websites. The scheme to obtain the interlinking networks, based on either the in-data set or the out-data set, then consists in identifying only the links between the websites of the sample within each data set (see Figure I). These interconnections between peers should form the same network from either the inlinks or outlinks. However, due to the differences in the gathering of the data sets and the inherent shortcomings and capabilities of the different sources, both data sets were compared in order to see the features of each one and determine how they could be used.

[**Please insert Figure 1 here**]

## Data analysis

### Structural analysis: comparing the IN & OUT data sets

The variety in the data set as well as the novelty of the approach applied to analyse the interactions calls for a comparison of the characteristics of the networks formed by the in- and out-data sets in order to gain empirical understanding of the holistic qualities of each data set. Hence, the structure of the networks are analysed with cohesion measures that help to estimate the degree of integration of the websites, the size of the network, the cohesion of the websites and the level of mutual interaction between them. *Inclusiveness* describes the number of websites who are integrated in the network and is the total number of nodes minus the isolated nodes (Scott, 2000). *Connectivity gap* is the total number of in- or outlinks established in both data sets minus the number of in- or outlinks in each data set and shows the proportion of links which the data set needs to reach the maximum number of potential links. *Density* refers to the extent to which all possible relations are actually present, and *reciprocity* indicates the proportion of relations (links) that are reciprocal (Wasserman & Faust, 1998).

**Table II. Structural cohesion measures of the in- and out-data sets for three SPs**

| 1) Advanced Manufacturing Park (AMP) | | | | | |
|---|---|---|---|---|---|
| | Inclusiveness | (%) | Ties | Connectivity Gap | Density | Reciprocity |
| in | 26 | (0.79) | 63 | 0.46 | 0.06 | 0.32 |
| out | 33 | (1) | 92 | 0.21 | 0.09 | 0.26 |
| both | 33 | (1) | 117 | | 0.11 | 0.36 |
| 2) Leeds Innovation Centre (LIC) | | | | | | |
| | Inclusiveness | (%) | Ties | Connectivity Gap | Density | Reciprocity |
| in | 45 | (0.92) | 97 | 0.21 | 0.04 | 0.29 |
| out | 49 | (1) | 106 | 0.14 | 0.05 | 0.28 |
| both | 49 | (1) | 123 | | 0.05 | 0.36 |
| 3) Yorkshire Science Park (YSP) | | | | | | |
| | Inclusiveness | (%) | Ties | Connectivity Gap | Density | Reciprocity |
| in | 103 | (0.99) | 307 | 0.19 | 0.03 | 0.33 |
| out | 104 | (1) | 312 | 0.17 | 0.03 | 0.37 |
| both | 104 | (1) | 378 | | 0.04 | 0.42 |

In the case of the *Advanced Manufacturing Park* (*AMP*), the comparison of the networks based on the in- and out-data sets shows that the interconnections based on the outlinks provide a larger relational structure which connects all the websites through a higher number and better distribution of the links. The in-network only integrates 80% of the websites and establishes 54% of the potential ties, forming a small and sparse network (see Table II). The cohesion measures based on the *Leeds Innovation Centre* (*LIC*) suggest that the in- and out-network are similar to each other, but it seems that the inter-inlinks can only build a smaller and more centralised network as a result of linking slightly less websites with a lower number of ties and a similar reciprocity. On the other hand, the outlinks integrate all the websites and form a slightly more dense network. In the case of *York Science Park* (*YSP*), again the differences between the in- and out-network are very small. The proportion of the connectivity gap, density and integration rate of both parks is almost the same, with the only difference being the higher degree of reciprocity and cohesion of the out data set.

From the initial overview of the three SPs, both data sets (inlinks and outlinks) provide similar structures and the patterns shown by the first SP differ from those of the second and third SP. This could be caused by the lower number of websites as well as the industrial profile of the AMP. Despite the fact that the in-data set is much larger than the out-data set, the latter might be the most effective because it seems to include more links among community members, hence forming the most complete community structure. It is also important to note that the inclusiveness of the out-data set has to be interpreted with caution because the websites of the sample were collected from the external outlinks of the SPs' websites, and thus all the websites should be linked. The high connectivity gap (0.23), the high rate of reciprocity when both data sets are combined, and the poor reliability of the webometric data suggest that the combination of both data sets could be necessary to ensure the most complete results.

In order to determine the degree of similarity and consistency of the interconnections established by the bi-directionality of the networks, Pearson correlations were calculated to measure the similarity between the in- and outlinks generated by the interlinking networks based on each data set (see Table III). The outlink counts obtained from the two data sets correlate significantly with (Pearson's $r$= 0.84, 0.98, 0.87), showing similar trends among the websites independently of the data set. In contrast, the correlation of the inlinks varies from SP to SP (Pearson's $r$= 0.58, 0.73, 0.94). The reason for the weak and strong correlation in the in- and outlinks of the *AMP* is the structural difference between both data sets. The in-data set of this SP is characterised by the low density caused by the few linked websites. However, the structural differences do not affect the outlinks to the same extent when the behaviour of the linking websites are compared because the role of the connectors in this type of network is monopolised by certain types of websites such as intermediaries, hybrid organisations and research institutions. Therefore, the high concentration of active websites facilitates the comparison between outlinks, and in contrast the wide dispersion of frequencies across the linked websites could make it difficult to find similarities when there are significant structural differences in both data sets.

**Table III. Pearson correlation between the in- and outlinks of each interlinking network**

|  | Inlinks | Outlinks |
|---|---|---|
| 1) Advanced Manufacturing Park (AMP) | 0.58 | 0.84 |
| 2) Leeds Innovation Centre (LIC) | 0.73 | 0.98 |
| 3) Yorkshire Science Park (YSP) | 0.94 | 0.87 |

The concentration of outlinks by a central group of websites that links a significant part of the network can be illustrated with the help of the Gini coefficient which measures the inequality of a distribution. It has previously been used to evaluate the web visibility of innovation systems (Katz & Cothey, 2006). The coefficient of 0 expresses total equality and a coefficient of 1 indicates maximal inequality. Hence, the inlinks should give a value closer to 0 whilst the outlinks with a higher concentration should give a value closer to 1. Table IV shows the inequality in the capacity to establish links in these networks and how the distribution followed by the in- and outlinks in both data sets is consistent. In the case of the *AMP*, the high coefficient of 0.74 for the inlinks in the in data set shows that only few websites are linked while the coefficient of 0.63 in the out-data set shows a slightly lower concentration in the distribution of the inlinks due to the higher number of interconnected websites. It helps to understand the structural difference of both data sets and the weak correlation of the inlinks for this SP. However, it is still necessary to study the local properties of the interlinking networks formed by both data sets and study the multilateral linkages.

**Table IV. Gini coefficients of the in- and outlinks of each data set**

|  | IN-data set | | OUT-data set | |
|---|---|---|---|---|
|  | Inlinks | Outlinks | Inlinks | Outlinks |
| 1) Advanced Manufacturing Park (AMP) | 0.74 | 0.85 | 0.63 | 0.86 |
| 2) Leeds Innovation Centre (LIC) | 0.38 | 0.85 | 0.38 | 0.84 |
| 3) Yorkshire Science Park (YSP) | 0.48 | 0.86 | 0.48 | 0.85 |

After a detailed analysis of the interconnections among the organisations of both structures, we found similar patterns of interaction, suggesting that both web data sets provide consistent structures. A visual overview of the structures (based on both data sets) is in Figure 2. Despite the consistency of the results, there are some connectivity gaps on either one or the other data set caused by a low overlap, and which may lead to the underrepresentation of certain weak ties and influence the visibility of certain types of organisations. This may result in loss of information and lead to misleading conclusions on a micro-level analysis. For example, we found that the in-data set provides better interconnectivity to academia, and the out-data set to government, while the links of industry are almost equally represented by both data sets. Therefore, the reliability problems of hyperlinks as an indicator and the inherent shortcomings of the data collection tools make it difficult to rely on only one data set for a webometric analysis since it can lead to inaccurate results (Thelwall, 2008a), especially in the study of dynamic and complex R&D networks where heterogeneous organisations converge and interact (Tijssen, 1998). Consequently, the combination of both data sets, collected from different sources, is recommended to obtain more robust and reliable interlinking networks.

Once the evidence indicates that the joint use of both dimensions provides more robust and reliable structures, the next step is to achieve the two objectives of this exploratory study. Subsequently, the combined networks were studied in order to find out first, whether these web-based R&D infrastructures are able to reflect the linkages across the institutional sectors and the different types of organisations that interact within the SPs, and second, whether the web-based patterns coincide with those found by indicators that measure the R&D activities to determine whether they are able to reflect some interesting offline features and thus answer the second goal of the study.

[*Please insert Figure 2 here*]

*Local analysis: combined interlinking networks*
1. *Interconnections between the institutional sectors and categories*

The number of links between the sectors, and the average of links by each organisation are reported in Table V. Here the outlinks of the SP's websites, which were used as seed URLs, are not taken into account to avoid redundant links. It shows that the academia sector, with the highest average out- and inlinks, is the most connected. The knowledge transfer function gives universities a central role, establishing links to public organisations that support the commercialization of academic knowledge and also to industry by means of firm-formation and consulting activities. The fact that academia links intensively but not in a reciprocal fashion is supported by previous findings (Garcia-Santiago & de Moya-Anegon, 2009). On the other hand, the high connectivity of government with itself and the low integration of industrial organisations also partly coincide and extend the findings of another webometric study (Stuart & Thelwall, 2006) where it was found that government is also well connected with itself, while only a little with the university and almost not connected with the industry. Nevertheless, a classification according to three institutional sectors may be too broad because some central organisations are the result of partnerships between two or the three sectors, being a combination of quasi-academic, -private or –public efforts.

**Table V. Interconnections between the institutional sectors**

| Websites | *Outlinks / Inlinks* | Industry | Academia | Government | *Total* | *Mean* |
|---|---|---|---|---|---|---|
| 122 | **Industry** | 26 | 13 | 43 | 82 | 0.67 |
| 12 | **Academia** | 26 | 20 | 45 | 91 | 7.58 |
| 52 | **Government** | 72 | 39 | 151 | 262 | 5.04 |
| 186 | *Total* | 124 | 72 | 239 | 435 | 2.34 |
| | *Mean* | 1.02 | 6.00 | 4.60 | 2.34 | |

In webometrics it is common that the number of links a website receives correlates with the size in the number of webpages (Aguillo et al., 2006; Ortega & Aguillo, 2008). Therefore, larger websites should obtain more links in the network. To observe if the size of the websites has a direct impact on the interlinking networks, Spearman correlations were calculated for the in- and out-degrees and the number of webpages of the organisations. The values (in-degree/webpages=0.46 and out-degree/webpages=0.46) show that there is a significant correlation and thus a tendency for larger web sites to be more central in the network. Therefore, the interconnectivity and position of the organisations within a R&D context may be partly influenced by their size on the web.

**Table VI. Top five organisations with the highest centrality measures for the three SPs**

| 1) Advanced Manufacturing Park (AMP) | | | | | |
|---|---|---|---|---|---|
| **Organisations** | **InDeg** | **Organisations** | **OutDeg** | **Organisations** | **Betweenness** |
| yorkshire-forward.com | 10 | attheamp.com | 31 | attheamp.com | 164.5 |
| ec.europa.eu | 8 | amptechnologycentre.co.uk | 20 | yorkshire-forward.com | 147.6 |
| twi.co.uk | 8 | yorkshire-forward.com | 14 | amrc.co.uk | 124.1 |
| amrc.co.uk | 7 | amrc.co.uk | 9 | amptechnologycentre.co.uk | 59.7 |
| materialise.com | 7 | ec.europa.eu | 7 | ec.europa.eu | 37.6 |
| **2) Leeds Innovation Centre** | | | | | |
| leeds.ac.uk | 13 | leedsinnovationcentre.com | 47 | leeds.ac.uk | 660.5 |
| yorkshire-forward.com | 6 | leeds.ac.uk | 18 | leedsinnovationcentre.com | 629.4 |
| hm-treasury.gov.uk | 6 | connectyorkshire.org | 9 | connectyorkshire.org | 189.4 |
| connectyorkshire.org | 6 | yorkshire-forward.com | 8 | yorkshire-forward.com | 121.3 |
| europa.eu | 5 | europa.eu | 4 | ukcrn.org.uk | 48.2 |
| **3) Yorkshire Science Park** | | | | | |
| york.ac.uk | 24 | york.ac.uk | 37 | york.ac.uk | 1467.7 |
| businesslink.gov.uk | 20 | sciencecityyork.org.uk | 31 | businesslink.gov.uk | 897.0 |
| yorksciencepark.co.uk | 18 | businesslink.gov.uk | 19 | sciencecityyork.org.uk | 835.3 |
| york.gov.uk | 13 | york-england.com | 19 | yorkshire-forward.com | 326.8 |
| sciencecityyork.org.uk | 12 | yorkshire-forward.com | 18 | york-england.com | 249.4 |

[*Please insert Figure 3 here*]

## 2. Advanced Manufacturing Park (AMP):

Located in Rotherham, the *AMP* is a joint venture between *Yorkshire Forward* and *UK Coal*. The park is designed to host manufacturing companies which specialise in precision manufacturing and advanced material technology processes (Advanced Manufacturing Park, 2010; Pullin, 2006). The AMP forms a network of 33 websites connected through 115 links, being the smallest of the three SPs (see Figure 3). The in- and the out-data sets provide 59 and 94 links respectively, and a combination of both gives an increase of 56 (93%) links for the in-data set and 21 (22%) for the out-data set. In the core of the network there are eight central organisations, including the regional development agency (RDA) *Yorkshire Forward,* with the highest in-degree and the third highest out-degree (see Table VI). Other important organisations are the SP (atteamp.com) and the *AMP Technology Centre*, which fills the role of an incubator. These intermediaries have the highest out-degrees, building a bridge between businesses and the core of the network. The main organisation responsible for the creation and diffusion of knowledge is the *Advanced Manufacturing Research Centre (AMRC),* a research centre established in partnership between the University of Sheffield, world-leaders in the aerospace supply chains, and government offices, and the *TWI Technology Centre* (twi.co.uk). The *AMRC* and *TWI* are two leading research organisations, which attract funding from contracts with public and private sector, and EU programmes (Hauser, 2010). The other three central support structure organisations are: *Business Link,* which delivers publicly funded business support products and services designed to help new businesses, the *EU-Regional Policy* (ec.europa), and the *Manufacturing Advisory Service* (mas-yh.co.uk) that delivers free and grant-funded advice as well as practical assistance to assist manufacturing businesses. These core organisations link important international giants and intensive R&D consultants. The central role of private and public sector funding in the development of the new ventures is illustrated by the collaboration between the business developer *LIFE-IC*, the RDA, and the incubator of the SP, with the spin-out from the *University of Leeds*, *Inertius*.

There are three features that should be noticed; first, the central position of the research center *AMRC* is caused by its strong formal ties with the consultants *TWI* and *Fripp Design*, and the business supporters *MAS* and *Business Link*. This sub-network is a collaboration that brings the research and technology to improve the competence of the local industry. Moreover, the *AMRC* is also supported by the RDA and the *EU Regional Development Fund (ERDF)*. Second, only the three consultants with most intensive R&D activities benefit from research contracts and alliances which bring together the knowledge, resources, expertise, and needs from three sectors. In contrast, the other consultants may be slightly isolated because they usually work in short term projects and commercialize a particular service, and thus do not require the same level of collaboration and resources (Bessant & Rush, 1995). Third, even though most of the businesses commercialise innovative products only eight firms are linked with government agencies, and four have direct links to technology producers and consultants.

[*Please insert Figure 4 here*]

## 3. Leeds Innovation Centre (LIC):

The *LIC* was established by the University of Leeds in 2000, and hosts a variety of companies, public organisations and university spin-offs and creates and attracts new healthcare and bioscience companies through its research, facilities and investments (Leeds Innovation Centre, 2010). *LIC* forms a network of 49 nodes connected trough 123 links, and the initial 90 links provided by the in-data set are increased in 33 (37%) while the 106 links of the out-data set constitute an increase of 17(16%). In the interlinking network three different subgroups can be identified: the peripheral subgroup on the left is basically formed by service-based firms and consultants, and by public organisations on the right, whilst the central subgroup is tied together by the academia, a transfer office, business developers and technology-based firms and university spin-offs. Hence, the interplay between the sectors occurs in the middle of the network and has the *University of Leeds* as the key connector. The other seven organisations in the core are: the RDA, three government institutions, the SP, and the business supporter *Investors in People*. The cross-fertilisation and knowledge transfer could be represented in the dynamic sub-network formed by the public seed capital facilitator, *Connect Yorkshire*, with the university which in turn ties the intellectual property and trademark company, *Techtran* (techtrangroup.com), as well as bio companies such as *Photopharmica*, and the university spin-outs *Instrumentel*, *LUTO Research*, *Chamelic* and *Tracsis*. Finally, it is noteworthy that the two nodes formed by the *Business Centre Association* (bca.uk.com) and the *University of Leeds Careers Centre* (careerweb.leeds.ac.uk) are less integrated than expected, since they both offer a range of services intended to help young entrepreneurs.

The R&D network is characterised by the leading position of the University and its ability to bridge the gap between the institutional spheres, connecting five spin-outs (instrumentel.com; tracsis.com; luto.co.uk; photopharmica.com; evidence.co.uk), business developers, support and public organisations, and health organisations. Being the first UK university to set up a dedicated technology transfer function (Lambert, 2003), the University's entrepreneurial identity is also materialized trough strong and formal ties with two consultants. In 2002 it became the first university in the UK to outsource its technology commercialisation activities to *Techtran* (IP Group, 2011). This partnership helps the university to identify the IP with high commercial potential and offers seed capital and strategic support services for new ventures. Consequently six of the nine knowledge-based firms identified in the park have spun out from the University of Leeds.

[*Please insert Figure 5 here*]

## 4. York Science Park (YSP):

*YSP* is a joint venture between the *University of York* and private investors and is situated on the campus of the University to stimulate the technology transfer with the knowledge-based enterprises located in the SP (UKSPA, 2010). Its tenants are related to bio- and health science and IT, having a similar profile to the SP of *LIC*. *YSP* forms the most heterogeneous and largest network with 104 websites and 378 links. The in- and the out-data sets provide 283 and 312 links respectively, and the combination of both supposes an increase of 95 (34%) links for the in- and 66 (21%) for the out-data set. Thus, in order to facilitate the representation and analysis of this comprehensive network, the implicit external links of the SP to all the websites were eliminated and only the incoming links of the SP were taken into account. It reduces the number of websites to 75 (72%) and the number of links to 275 (73%). In the new network, it is found three brokers that link different sub-networks together (see Figure 5). The most important broker is the *University of York* that intermediates between a group of knowledge-based firms and spin-outs, and various consultancy offices, public and non-government organisations with the central support structure organisations and business developers. The second broker is the business developer *Science City York* (*SCY*) that builds a bridge between the firms with the University and the central R&D structure. The third broker is the SP which is connected by firms, and organisations related to the movement of the SPs in the UK as well as support organisations. The number of links received by the SP is also significant due to the fact that this differs from the patterns showed by the previous SPs. The other eight organisations in the core of the network are: the support structure organisations and local authorities such as *Business Link*, the RDA, *York-England* and *City of York Council* which attract new investments and provide economical resources, specialised advice and support for the network. These supporters allow the establishment of partnerships and delivery of funds oriented to other central intermediaries and networking organisations such as *Higher York* that is a partnership between local Higher Education Institutions to offer professional training, the local *Chamber of Commerce*, *Business Link Yorkshire*, and *York Professionals*. The tight intersection of the three sectors act as drivers for the whole network and are in turn linked to close partners. The central position of the intermediaries and support organisations reflects the importance of these hybrid nodes in the development of the whole network and its fundamental role in innovative contexts.

The supportive infrastructure is again tied together by the active role of the University in partnership with the other institutional sectors. This collaboration facilitates the commercialisation of academic knowledge and technology through consultancy and firm-formation activities which are in line with university's emerging social and entrepreneurial mission. The collaboration between the University and support structure organisations, especially the RDA, has set up the seed capital investor *White Rose Technology Seedcorn Fund* (whiteroseseedcorn.com), and business developer *SCY*, and has also supported the investor *Yorkshire Association of Business Angels* (yaba.org.uk), which together promote the creation and growth of business through public supported mentoring and facilitating investment funds that has an significant impact on the local fast growing knowledge-based industry (Lambert, 2003). This R&D structure may be the main reason that nine university spin-offs are based on the SP. There are seven from the *University of York* (yorkmetrics.com; cell-analysis.com; origin-consulting.com; rapitasystems.com; xceleron.co.uk; pro-curetherapeutics.com; cybula.com) and two attracted from *Leeds* (avacta.com; tissueregenix.com). The similar number of service-oriented and knowledge-based firms in the SP could be the result of a wide range of firms that are supported by the *SCY*, which supports both type of firms. Finally, the consultancy services offered by the University could be dived into internal activities, through the research contracts with independent bodies and NGOs, and external activities, through quasi-academic businesses.

# Discussion

The representation of the links established between the various types of organisations provides a fast and broad overview of the articulation of the R&D networks, identifying the interactions and key organisations which could be expected to be found in a highly institutionalised environment such as organizational innovation structures. The analysis of the local properties of the interlinking networks shows that the structures of the SPs tend to be organised according to the academic, industrial and governmental sectors. The degree of interaction between the groups makes it possible to observe that some of the firms and some governmental organisations tend to occupy peripheral positions. These two groups are tied together by the third group which occupies the centre of the network. The central sub-network is led by academia and support structure organisations which support the rest of the network. In this overlapping area the groups of support and knowledge generation as well as the knowledge exploitation interplay in order to foster an innovative environment. The close interactions among research organisations, which form the knowledge generation subsystem, and how this group is linked by the companies, which constitute the knowledge exploitation subsystem, reflects the importance of the interdependence and interactive learning between the two subsystems in innovative contexts (Coenen, et al., 2004). The cross-sectoral interactions are also facilitated by hybrid organisations, such as the business developers, which act as the main interface with the private sector and promote the exploitation of academic knowledge trough the creation and support of a knowledge-based industry. The leadership of the universities and research centres confirm the central role of the academia as a principal source of innovation in the knowledge-based economy (Etzkowitz, 2008; Etzkowitz & Leydesdorff, 2000), and especially in this infrastructures where, as expected, the SPs are used as platforms to foster collaboration and cross-fertilisation. On the other hand, the SPs also play an important role in the network, connecting the businesses with the rest of the central organisations.

The analysis of the three SPs uncovers interesting patterns which help to gain a first insight into how the organisations in these innovative social circles engage on the web, and how these links may be proofs of offline relationships. In order to answer the second goal of this study, it is necessary to find out whether the key features observed in the three web-based R&D networks correspond to the actual features of R&D infrastructure developed in the region of Yorkshire and the Humber. For this purpose, reports and surveys commissioned and produced by the Higher Education Funding Council for England (HEFCE) and the Department for Business, Innovation and Skills (BIS) have been consulted and used for the comparative analysis.

In summary, despite the fact that the similar structures formed by *LIC* and *YSP* differ from the small network formed by the industrial *AMP*, there are some interesting shared properties, such as the prominent role of the research producers, the public support and commitment, the transference of knowledge fostered by the networking capabilities of business developers, and the knowledge exploitation through contract research, consultancy services and creation of new ventures. In the *AMP* the research centre *AMRC* is the anchor tenant and forms the core of the network, showing how this flexible and integrated research facility works in collaboration with government, academia and industry to provide the network with the advanced technology to compete at an international level (Technology Strategy Board, 2010). In *LYC* the *University of Leeds* emerges as the key connector and a local innovation organizer, using the SP as the ideal quasi-academic platform to commercialize its academic research and technology through knowledge transference, firm formation and consultation activities (Lambert, 2003). Unlike the *LYC*, the *University of York* not only uses the SP to interact more easily with industry and government and nurture a dynamic environment. *YSP* is also used a tool to embrace both an active academic entrepreneurial activity to increase the competitive advantage of the local industry as well as a wider social mission designed to collaborate and support the civil society, as indicated by the number of public organisations and NGOs based on the park. As Etzkowitz (2006) argues "the [York] University has moved into a more central role both in the York region and in the larger society as a generator of new business firms... These developments have led to a new format of academic and business organisation, bringing together elements of each at a common site. ... It has successfully replaced declining industries and is now positioning itself to become the core of the city-region's economy".

Second, the efforts of the local and regional authorities to restructure the traditional industry into a knowledge-based one could be reflected by the central position of support structure organisations such as RDA, *Business Link*, RDEF, whose aim is to support the infrastructure as strategic drivers of regional development (Hauser, 2010). This hands-on approach supposes the delivery of the highest public investments in the UK to develop a R&D infrastructure (BIS, 2009), and goes hand in hand with one of the highest academic spending in R&D (Lambert, 2003) with the collaboration between government and academia as the main driving force behind the networks. This structural collaboration has led to design networks which are able to provide the capital, advice, and resources through early-stage investment facilitators and business developers, fostering a growing knowledge-based industry. Consequently the networking capabilities and hybrid roots of business developers stimulate higher cohesion between the institutional spheres. In the case of *LIC* and *YSP*, both have different knowledge transfer models. *LIC* takes advantage of the University's strong links with a private IP office and focuses on licensing activities and supporting the formation of promising spinouts, while *YSP* has a public-university oriented approach that focuses on setting up a wide range of businesses (Lambert, 2003). This difference in both approaches can be identified by the selected number of knowledge-intensive spin-offs that have flourished despite the limited presence of business supporters in *LIC*, while *YSP* promotes the development of a wider range of knowledge-based and services-oriented firms which are supported by a larger public infrastructure. Hence, there is a system that only invests in deals which suppose low-risk and high returns, and another that invests to obtain immediate results in the form of firm and job creation, independently of the quality and sustainability of the ventures in the long term (Nottingham BioCity, 2010). Apart from that, the role of publicly backed investments in the establishment of business developers and in the provision of redirected funds observed in the web-based networks may match the increasingly representation of the public sector in the capital market, where participated in 42% of all venture capital deals in 2009 and 68% of all early-stage deals in 2008 in the UK (Pierrakis, 2010).

Fourth, it seems that only most research intensive consultants are able to link research producers and other organisations. In the case of the *AMP*, the high number of R&D consultants and the pivotal role of the *AMRC* in the diffusion and application of new advanced technology show that the SP functions as a seedbed for the industry in joining and engineering development (House of Commons, 2011; Technology Strategy Board, 2010). However, only most R&D intensive consultants and research consortiums are able to serve as liaisons for the institutional spheres and develop technology transfer programmes (Bessant & Rush, 1995). This fact is observed in the *AMP* and *YSP* where the consultants with a research basis and academic roots are involved in a more complex and institutionalized process, obtaining a better integration. Although, the weak connectivity of the consultants may depend on the reticence to make public commercial relationships or on establishing the links with clients outside the SP, which makes it necessary to study the external visibility of these organisations to determine the value of their services in the national and international arena. Here it is worth noting that the incomes of the academic sector, regarding business and community interaction, are mainly obtained trough non-commercial research contracts whereas consultancy, IP, and the use of facilities and equipment provide marginal incomes, as well as the public and third sector organisations are the major clients and provide the highest incomes in the region (HEFCE, 2010). Consequently, the interactions identified in the form of research contracts between the *University of York* with public and third sector organisation based on *YSP* proved to be representative.

Finally, the university spin-offs are the other identified forms of knowledge exploitation. As expected, in these structures the considerable number of spin-offs illustrates a strong research basis and the conditions achieved by the R&D infrastructures. However, although the collaborative process to establish and develop knowledge-based firms leads to a better integration than service-oriented firms, the private sector still occupies a peripheral position. This may be caused by the lack of private investments in the region (Nottingham BioCity, 2010) and the lack of linkages among firms located in SPs (Quintas et al., 1992; Suvinen, Konttinen, & Nieminen, 2010), as well as the inaccuracy of direct links to detect commercial ties (Stuart & Thelwall, 2006; Vaughan & You, 2006). Nevertheless, we should be aware that the proportion of collaboration among firms engaging in innovation activities is only of 23%, and the three less frequent partners as well as sources of innovation information are furthermore consultants and private R&D institutes, Universities, and Government or public research institutes, respectively (BIS, 2009). Therefore, regardless the application of direct or indirect links it could still be difficult to find signs of cooperation, also among innovating enterprises.

## Conclusions

This exploratory study introduced a web-based approach based on interlinking networks within a community. Given that the interconnections within a group of websites should derive the same structure regardless of the use of in-links or out-links, both dimensions were collected from three different sources and afterwards, used jointly in an attempt to obtain a more robust and reliable structure. A structural comparison confirmed that both dimensions provide similar structures, because there was a significant correlation between the links generated by both data sets. We also found that the out-dimension, collected from Bing and SocSciBot, with half of the total links of the in-dimension, collected from Yahoo!, provided more cohesive structures. In addition, the overlap between Bing and the crawler SocSciBot is only of 4%.

In answer to the first research question: Can webometrics methods identify interactions between institutional sectors as well as interactions between various types of organisations associated with SPs? The combined interlinking networks show that the SPs tend to be organised according to the academic, industrial and governmental sectors, and that they are primarily interconnected by a central sub-network which is led by academia and support structure organisations which in turn support the rest of the network. These web-based networks probably reflect collaboration between academia and government to generate an infrastructure that hosts heterogeneous organisations which are dedicated to exploiting the local research basis in order to promote knowledge-intensive industry in the region. In answer to the second research question: Are the main features of SP interlinking networks in line with findings of official reports and surveys which are essential sources of information to evaluate the R&D infrastructure in the UK? The analysis of the main features shared by the web-based R&D networks seems to reflect the potential strengths and weaknesses of the policy measures and conditions described by different surveys and reports on the R&D infrastructure developed in the region. According to the evidence based on traditional indicators the web-based network may reflect the prominent role of the research producers and the commitment of the public support through various support structure organizations. Both sectors make significant economic efforts and closely interact to foster networks that are able to provide the capital, advice, and resources. The difference in the private and university-public knowledge transfer models followed in the SPs may also be illustrated by the type of firms that are established and by the size of the support infrastructures developed. In addition, the significant presence of the public sector identified also matches its representation in the capital market. The knowledge exploitation mechanisms observed are contract research, consultancy services and creation of new ventures. The evidence found seems to confirm the fact that only the most research intensive consultants are able to interact with research producers and other organisations, and that the universities' research contracts are primarily obtained from public and third sector organizations. The considerable number of spin-offs indicates a strong research basis and the conditions achieved by the R&D infrastructures, while its low degree of linkage among firms located in SPs corroborates previous studies.

The novel method introduced is able to extend previous webometric attempts that study R&D networks and provide a broad overview of a complex and highly institutionalised innovation infrastructures developed in the studied region and that could be further analysed by other approaches. This web-based approach may facilitate the study of a complex underlying structure that needs to be assessed by studies that focus on the different aspects and particular actors that are embedded in these technology and innovation centres. The first findings suggest that it may be useful to investigate the social and entrepreneurial activities of the university, large technology and innovation centres, clusters, regional innovation strategies, and dynamic systems with a high institutional heterogeneity. However, due to the exploratory nature of this paper and the early stage of webometric studies, these findings are merely indicative and additional research is required. A clear limitation of this approach is only taking into account the hyperlinks between the websites of the sample and not using information about specific inter-organization connection in the SPs studied. This means that the bias introduced by the use of external connections through spam and irrelevant connections is ignored, but at the same time these internal links may not reflect all the potential relations established so the results should be taken as illustrative rather than exhaustive. An example of this limitation is the *AMP*, that represents a successful model that has been exported to other 9 countries (House of Commons, 2011; Technology Strategy Board, 2010) but the low number of links with industry found may suggest that the analysis is only able to uncover a small part of the technology transfer and innovation process which takes place in Sheffield. Consequently, questions about the role of the *University of Sheffield* and about the impact of the research centre and consortiums outside the SP remain unanswered. The low visibility of organisations that establish commercial ties, like firms and consultants, may be also affected by the use of direct links, suggesting the use of indirect links could be also necessary. Finally, it is worth noting that the analysis of various types of organisations may lead to a bias towards the public and academic organisations which tend to have large websites and then be more visible, and also that the traditional classification (industry-university-government) is not always appropriate in R&D networks, because it is too broad to show the hybridization process of the institutional sectors to establish some important central organisations. Future studies should therefore evaluate the R&D networks constructed based on direct and indirect links to determine the potential similarities and differences of both approaches, design a framework to identify and evaluate the behaviour of the institutional agents on the web, and design a method to identify the interactions that could be relevant in the external environment of the SPs.

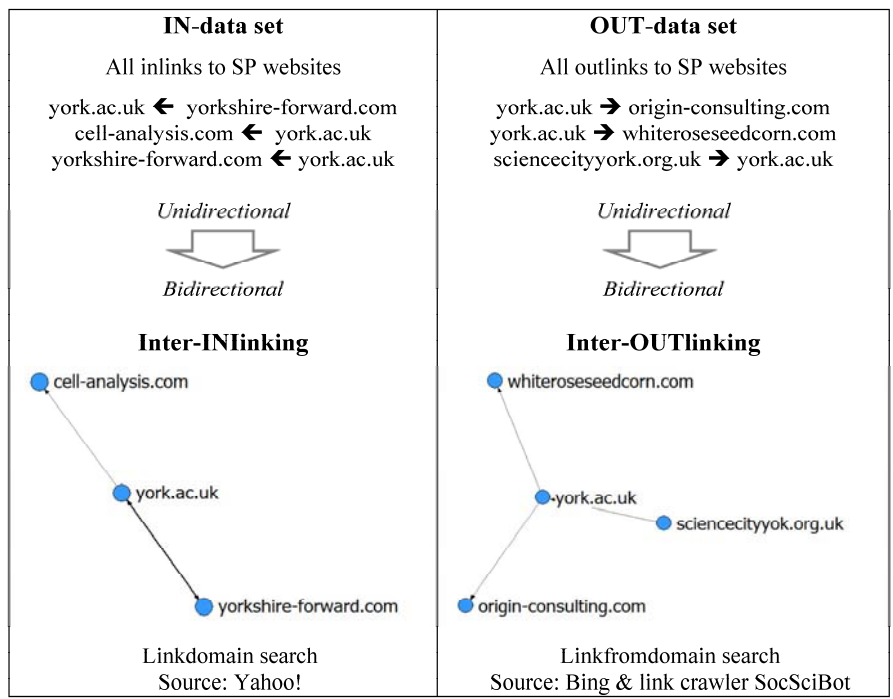

**Figure 1. Creation of interlinking networks**

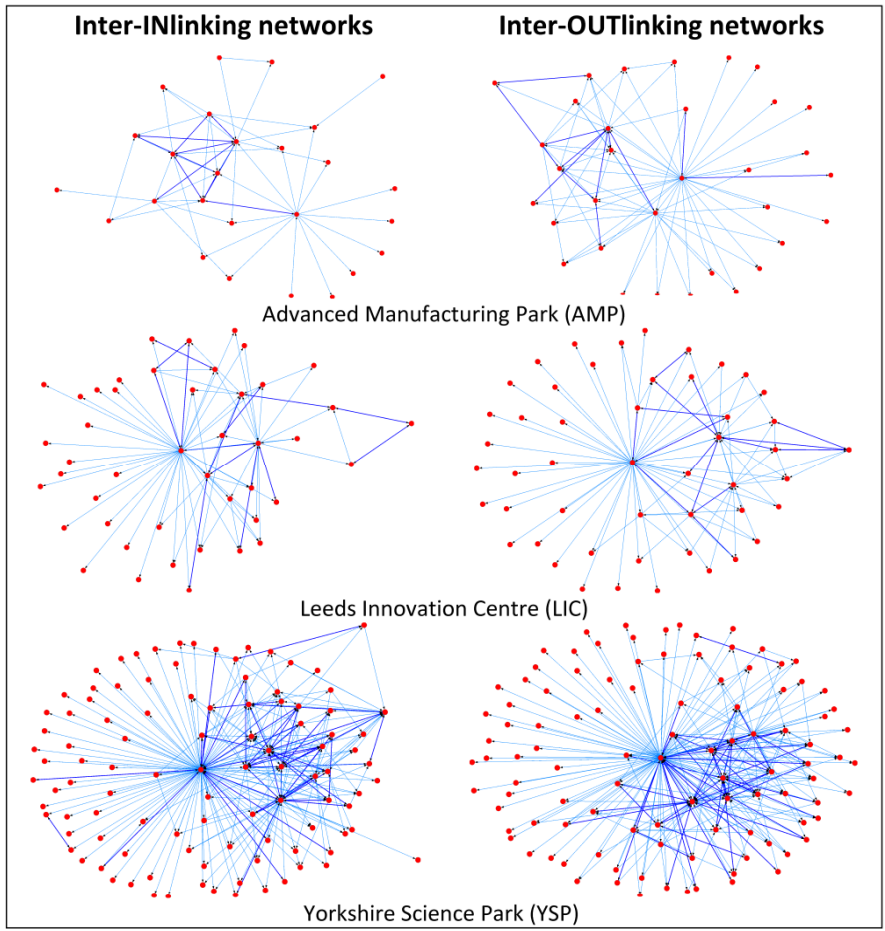

**Figure 2. Visual overview of the structures based on both data sets. The thicker lines indicate mutual linkages**

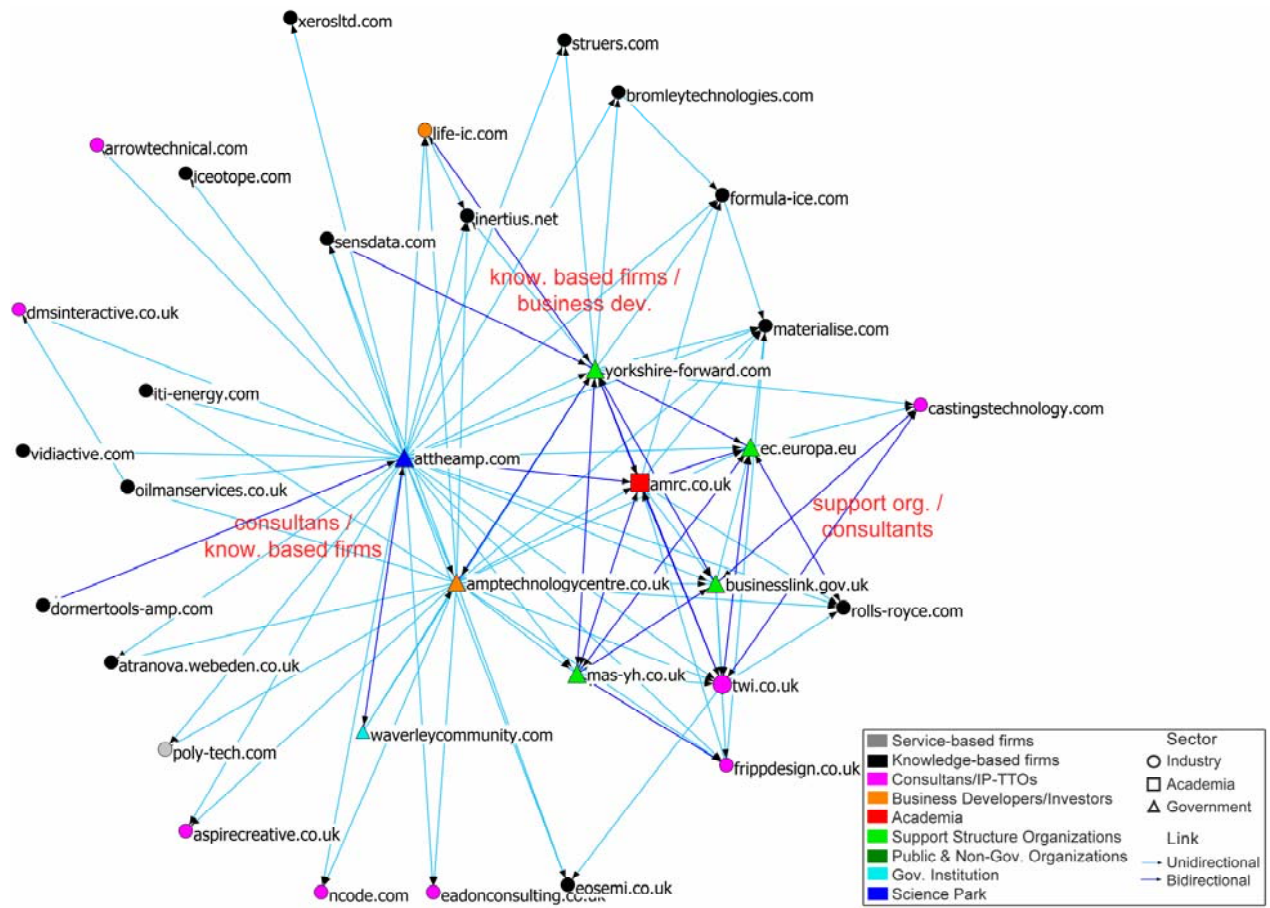

**Figure 3.** Inter-inlinking network of the Advanced Manufacturing Park (AMP) (*A colour version is available at: http://home.wlv.ac.uk/~in1493/11/fig-3.jpg*)

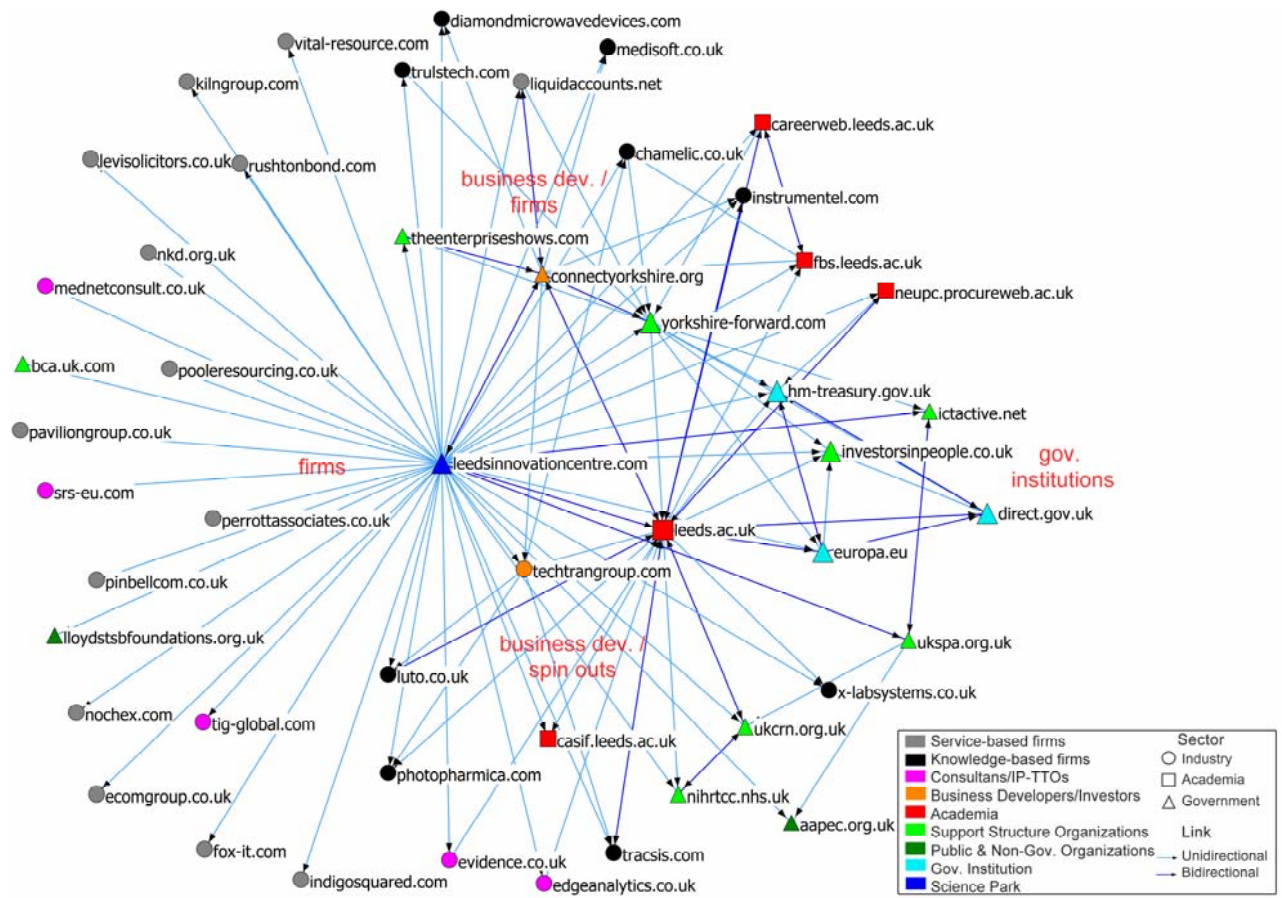

**Figure 4. Inter-inlinking network of the Leeds Innovation Centre (LYC) (A colour version is available at: http://home.wlv.ac.uk/~in1493/11/fig-4.jpg)**

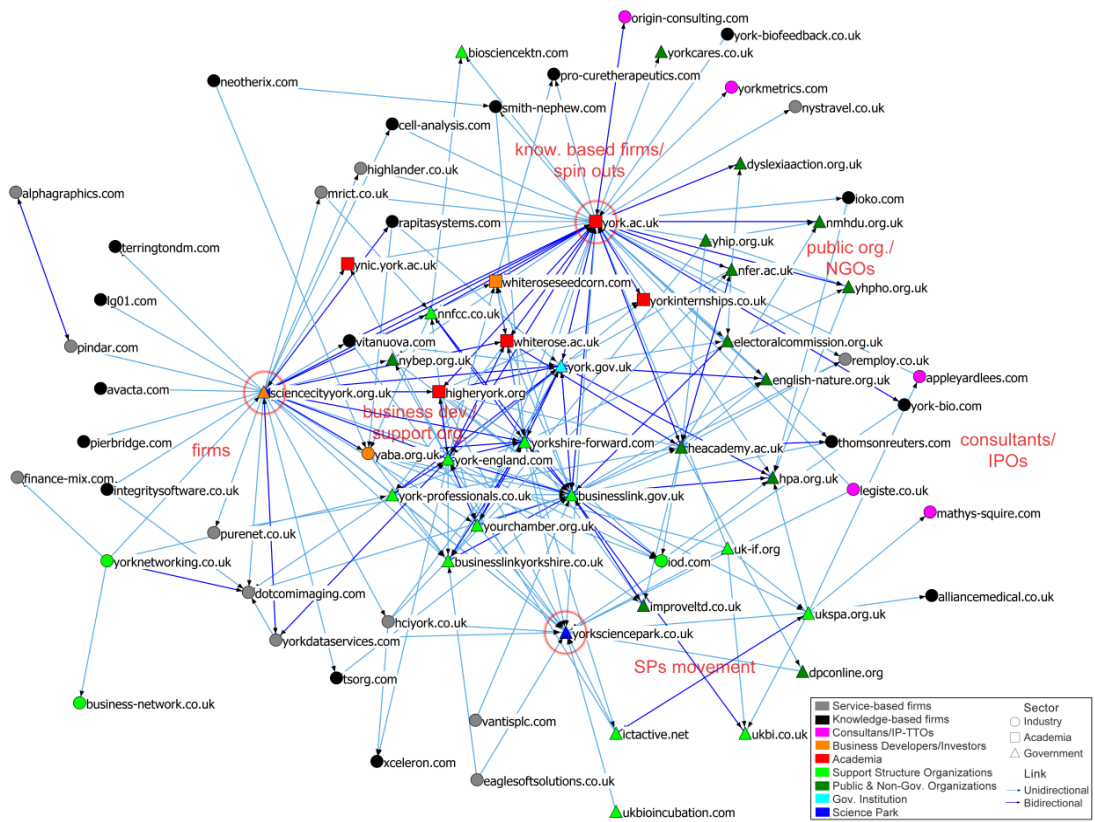

**Figure 5. Inter-inlinking network of the York Science Park (YSP) (A colour version is available at: http://home.wlv.ac.uk/~in1493/11/fig-5.jpg)**